\documentclass[pdflatex,sn-mathphys-num]{sn-jnl}

\usepackage{graphicx}%
\usepackage{multirow}%
\usepackage{amsmath,amssymb,amsfonts}%
\usepackage{amsthm}%
\usepackage{mathrsfs}%
\usepackage[title]{appendix}%
\usepackage{xcolor}%
\usepackage{textcomp}%
\usepackage{manyfoot}%
\usepackage{booktabs}%
\usepackage{algorithm}%
\usepackage{algorithmicx}%
\usepackage{algpseudocode}%
\usepackage{listings}%

\begin{document}

\title[Article Title]{Wave Turbulence and Cortical Dynamics}

\author[1,2]{\fnm{Gerald K.} \sur{Cooray}}\email{gerald.cooray@ki.se}

\affil[1]{\orgdiv{Clinical Neuroscience}, \orgname{Karolinska Institutet}, \orgaddress{\street{Eugeniav}, \city{Stockholm}, \postcode{17177}, \country{Sweden}}}
\affil[2]{\orgdiv{GOS-UCL}, \orgname{University College of London}, \orgaddress{\street{30 Guilford St}, \city{London}, \postcode{WC1N 1EH}, \country{UK}}}


\abstract{Cortical activity recorded through EEG and MEG reflects complex dynamics that span multiple temporal and spatial scales. Spectral analyses of these signals consistently reveal power-law behaviour, a hallmark of turbulent systems. In this paper, we derive a kinetic equation for neural field activity based on wave turbulence theory, highlighting how quantities such as energy and pseudo-particle density flow through wave-space ($k$-space) via direct and inverse cascades. We explore how different forms of nonlinearity—particularly 3-wave and 4-wave interactions—shape spectral features, including harmonic generation, spectral dispersion, and transient dynamics. While the observed power-law decays in empirical data are broadly consistent with turbulent cascades, variations across studies—such as the presence of dual decay rates or harmonic structures—point to a diversity of underlying mechanisms. We argue that although no single model fully explains all spectral observations, key constraints emerge: namely, that cortical dynamics exhibit features consistent with turbulent wave systems involving both single and dual cascades and a mixture of 3- and 4-wave interactions. This turbulence-based framework offers a principled and unifying approach to interpreting large-scale brain activity, including state transitions and seizure dynamics.}




\maketitle

\section{Introduction}

Cortical activity at small spatial and temporal scales often exhibits oscillatory transients with a complex and variable mixture of frequencies \cite{buzsaki2023brain,buzsaki2006rhythms}. This activity has been effectively modelled using nodal approaches, where each node follows neural mass dynamics with a complex interaction between nodes. Neural mass models are widely used to simulate brain signals such as EEG and MEG by modelling interactions between excitatory and inhibitory neural populations \cite{david2003neural,byrne2020next,jansen1995electroencephalogram}.
At a more mesoscopic level, the cortex, with its convoluted sheet-like geometry, is well represented as a field defined over a two-dimensional surface. Many models of cortical dynamics describe propagating waves of activity, often interacting non-linearly across both space and time—or equivalently, across time and frequency domains \cite{cook2022neural,coombes2005waves,lopes1974model,bressloff2011spatiotemporal,wilson1973mathematical}.

Several empirical features of cortical activity constrain the development of such models, including signal decay and finite wave transmission speeds. Despite these constraints, the dynamics remain highly complex, as evidenced in electroencephalogram (EEG) recordings, particularly in subjects with neurological disorders such as epilepsy \cite{jirsa2014nature,lagarde2019repertoire}.

A prominent and consistent finding in recordings of cortical activity—whether at the level of local field potentials (LFPs) from neural populations, intracranial electrodes sampling from regions of approximately $1mm^3$, or scalp electrodes measuring from several $cm^2$—is the presence of a spectral power-law decay in signal power across frequencies \cite{he2014scale}. This $\frac{1}{f^a}$ scaling is a hallmark of turbulent flow dynamics \cite{donoghue2022methodological}.

Turbulence has been extensively studied in fluid dynamics, with a rich literature describing both experimental observations and rigorous theoretical frameworks derived from fundamental equations such as the Navier–Stokes equation. This non-linear partial differential equation exhibits turbulent solutions under certain conditions, where perturbed equilibria evolve into steady states characterised by a constant flux of energy or other conserved quantities. The mathematical description of such turbulent states was pioneered by Richardson and Kolmogorov \cite{kolmogorov1991local,frisch1995turbulence}, whose work accurately predicted the spectral features of turbulent energy cascades.

It is striking that neural field dynamics—despite their differing physical foundations—also exhibit spectral features reminiscent of those seen in turbulence. This resemblance can be more naturally understood in the framework of wave turbulence (or weak wave turbulence), a well-developed theory describing non-linear interactions between waves that generate turbulent-like spectral distributions \cite{nazarenko2011wave,zakharov2012kolmogorov}.

Recent work has proposed that cortical dynamics may indeed follow principles of turbulent wave activity, accounting for the observed power-law behaviour in cortical spectra \cite{deco2020turbulent,deco2025turbulence}. Such turbulence may influence many aspects of brain function, including the rapid transitions between different oscillatory states. In particular, epileptic seizures—sudden and dramatic disruptions of cortical dynamics are accompanied by characteristic time–frequency changes. These seizure-associated dynamics will also be analysed and modelled in this work to assess whether turbulent wave frameworks can capture their essential features.

In this paper, we begin with a model of neural activity and derive from it the characteristic behaviour of turbulent flow. We then compare these theoretical predictions with experimental results from cortical recordings. We show that many empirical findings can be plausibly explained by assuming that cortical activity is governed by wave turbulence.

\section{Model}
We begin with the cortical dynamics previously analysed in several studies \cite{cooray2023global,cooray2023NF,cooray2024cortical}. The evolution of cortical activity is governed by the following equation:
\begin{equation}
    \begin{aligned}
        i\partial_{t}\phi(r) & = \mathcal{F}(\phi)\\
     \end{aligned}
\end{equation}
The neural field variable, $\phi$, is a complex scalar where the real and imaginary components reflect activity in excitatory and inhibitory layers of the cortex. The functional $F$ maps the neural field activity across all spatial points to a complex scalar that describes the change at a given point. This represents the most general form used to characterise cortical field dynamics. To establish the starting point for our analysis, we express the dynamics in terms of the Fourier transform of the field $\phi$.

\begin{equation}
    \begin{aligned}
    \int \phi(\mathbf{r})e^{-i\mathbf{k}.\mathbf{r}} & = \psi(\mathbf{k})\\
    i\partial_{t}\psi(\mathbf{k}) & = \mathcal{G}(\psi)\\
      \end{aligned}
\end{equation}
The system can be studied analytically if it is integrable, meaning that the functional can be reformulated in terms of a Hamiltonian. Noting that $\psi$ is complex valued, the dynamics can be written as follows:
\begin{equation}
    \begin{aligned}
    i\partial_{t}\psi(\mathbf{k}) & = \frac{\delta \mathcal{H}(\psi)}{\delta \psi^*}\\
      \end{aligned}
\end{equation}
The linearised form of the cortical field dynamics has been shown to follow wave equations \cite{cooray2023NF}. We begin by expanding the functional $\mathcal{H}$.
\begin{equation}
    \begin{aligned}
    \mathcal{H} & = \mathcal{H}_0 + \mathcal{H}_{int}\\
      \end{aligned}
\end{equation}
This set of equations allows us to analyse the dynamics using perturbation theory around the solution corresponding to $\mathcal{H}_0$. A Hamiltonian such as $\mathcal{H}_0$, which leads to wave equations, can often be written, after a suitable canonical transformation of the $\psi$ variables, in a simplified form amenable to analysis.
\begin{equation}
    \begin{aligned}
    \mathcal{H}_0 & = \int d\mathbf{k}G_2(\mathbf{k})\psi\psi^*\\
      \end{aligned}
\end{equation}
We then get the linearised solutions.
\begin{equation}
    \begin{aligned}
    i\partial_{t}\psi(\mathbf{k}) & = \omega(\mathbf{k})\psi\\
    \psi & = Ae^{-i\omega(\mathbf{k})t}\\
    \phi & = A'e^{-i(\omega(\mathbf{k})t\pm \mathbf{k}.\mathbf{r})}\\
      \end{aligned}
\end{equation}
If the $\omega$-term is independent of $\mathbf{k}$ we get the Klein-Gordon equation which was derived in \cite{cooray2023NF}. We will not assume this in this study but allow for the following variation.
\begin{equation}
    \begin{aligned}
    \omega(\mathbf{k}) & = C|\mathbf{k}|^\alpha\\
    \end{aligned}
\end{equation}
There is empirical evidence suggesting that the value of $\alpha$ is greater than 1, indicating decaying waves. This leads to dispersion of wave packets on the cortical surface, as higher-frequency components travel faster than lower-frequency ones, resulting in attenuation and spreading of the wave packet over time. The interaction term $\mathcal{H}_{\text{int}}$ introduces non-linear coupling between waves. It is possible to derive kinetic equations that describe the flow of waves in $k$-space. By expanding the interaction functional in terms of contributions involving an increasing number of interacting waves, we obtain a series of interaction terms. In this work, we focus on three-wave and four-wave interactions. It has been shown that the dominant dynamics are typically captured by the lowest-order non-vanishing interaction term. For a detailed derivation of the kinetic equations, we refer the reader to the following standard texts: \cite{zakharov2012kolmogorov,nazarenko2011wave}.

 \begin{equation}
    \begin{aligned}
    \mathcal{H}_{int} & = \int d\mathbf{k}_1d\mathbf{k}_2d\mathbf{k}_3\bigg[G_{13}(\mathbf{k}_1,\mathbf{k}_2,\mathbf{k}_3)\left(\psi(\mathbf{k}_1)\psi(\mathbf{k}_2)\psi(\mathbf{k}_3)\delta(\mathbf{k}_1+\mathbf{k}_2+\mathbf{k}_3)+c.c\right)\\
&+G_{23}(\mathbf{k}_1,\mathbf{k}_2,\mathbf{k}_3)\left(\psi(\mathbf{k}_1)\psi(\mathbf{k}_2)\psi^*(\mathbf{k}_3)\delta(\mathbf{k}_1+\mathbf{k}_2-\mathbf{k}_3)+c.c\right)\bigg]\\
&\int d\mathbf{k}_1d\mathbf{k}_2d \mathbf{k}_3\mathbf{k}_4  \bigg[G_{14}(\mathbf{k}_1,\mathbf{k}_2,\mathbf{k}_3,\mathbf{k}_4)\left(\psi(\mathbf{k}_1)\psi(\mathbf{k}_2)\psi(\mathbf{k}_3)\psi(\mathbf{k}_4)\delta(\mathbf{k}_1+\mathbf{k}_2+\mathbf{k}_3+\mathbf{k}_4)+c.c\right)\\
&+G_{24}(\mathbf{k}_1,\mathbf{k}_2,\mathbf{k}_3,\mathbf{k}_4)\left(\psi(\mathbf{k}_1)\psi(\mathbf{k}_2)\psi^*(\mathbf{k}_3)\psi^*(\mathbf{k}_4)\delta(\mathbf{k}_1+\mathbf{k}_2-\mathbf{k}_3-\mathbf{k}_4)+c.c\right)\\
&+G_{24}(\mathbf{k}_1,\mathbf{k}_2,\mathbf{k}_3,\mathbf{k}_4)\left(\psi(\mathbf{k}_1)\psi(\mathbf{k}_2)\psi(\mathbf{k}_3)\psi^*(\mathbf{k}_4)\delta(\mathbf{k}_1+\mathbf{k}_2+\mathbf{k}_3-\mathbf{k}_4)+c.c\right)\bigg]\\
      \end{aligned}
\end{equation}
Introducing some shorthand notation we can write the above as.
 \begin{equation}
    \begin{aligned}
    \mathcal{H}_{int} & = \int d\mathbf{k}_1d\mathbf{k}_2d\mathbf{k}_3\bigg[G^{13}_{123}\left(\psi_1\psi_2\psi_3\delta(\mathbf{k}_1+\mathbf{k}_2+\mathbf{k}_3)+c.c\right)\\
&+G^{23}_{123}\left(\psi_1\psi_2\psi_3^*\delta(\mathbf{k}_1+\mathbf{k}_2-\mathbf{k}_3)+c.c\right)\bigg]\\
&\int d\mathbf{k}_1d\mathbf{k}_2d \mathbf{k}_3\mathbf{k}_4  \bigg[G^{14}_{1234}\left(\psi_1\psi_2\psi_3\psi_4\delta(\mathbf{k}_1+\mathbf{k}_2+\mathbf{k}_3+\mathbf{k}_4)+c.c\right)\\
&+G^{24}_{1234}\left(\psi_1\psi_2\psi^*_3\psi^*_4\delta(\mathbf{k}_1+\mathbf{k}_2-\mathbf{k}_3-\mathbf{k}_4)+c.c\right)\\
&+G^{24}_{1234}\left(\psi_1\psi_2\psi_3\psi^*_4\delta(\mathbf{k}_1+\mathbf{k}_2+\mathbf{k}_3-\mathbf{k}_4)+c.c\right)\bigg]\\
      \end{aligned}
\end{equation}
The different terms of the Hamiltonian function describe different types of wave interactions. The first term describes 3 wave interactions. These waves are valid only for waves with a decay in energy, i.e. if $\alpha>1$.
 \begin{equation}
    \begin{aligned}
    \mathcal{H}_{3} & = \int d\mathbf{k}_1d\mathbf{k}_2d\mathbf{k}_3\bigg[G^{13}_{123}\left(\psi_1\psi_2\psi_3\delta(\mathbf{k}_1+\mathbf{k}_2+\mathbf{k}_3)+c.c\right)\\
&+G^{23}_{123}\left(\psi_1\psi_2\psi_3^*\delta(\mathbf{k}_1+\mathbf{k}_2-\mathbf{k}_3)+c.c\right)\bigg]\\
    \end{aligned}
\end{equation}
The first term will only be possible with negative energy waves as the process described annihilation or creation of waves, and assuming that this is not allowed in the neural field dynamics we are left with the last term. This process will be a $2 \rightarrow 1$ or a $1 \rightarrow 2$ wave process. Later on we will need to investigate a statistical collection of neural fields where averaging over the phase variable is required. The above processes of unequal number of waves before and after the interaction will lead to non-zero contributions if paired with the appropriate conjugate term. This would be due to higher order pertubation terms from the  $\mathcal{H}_{3}$ interaction, Figure 1. We have for the 3-wave hamiltonian the following expression, 

\begin{figure}
    \centering
    \includegraphics[width=1\linewidth]{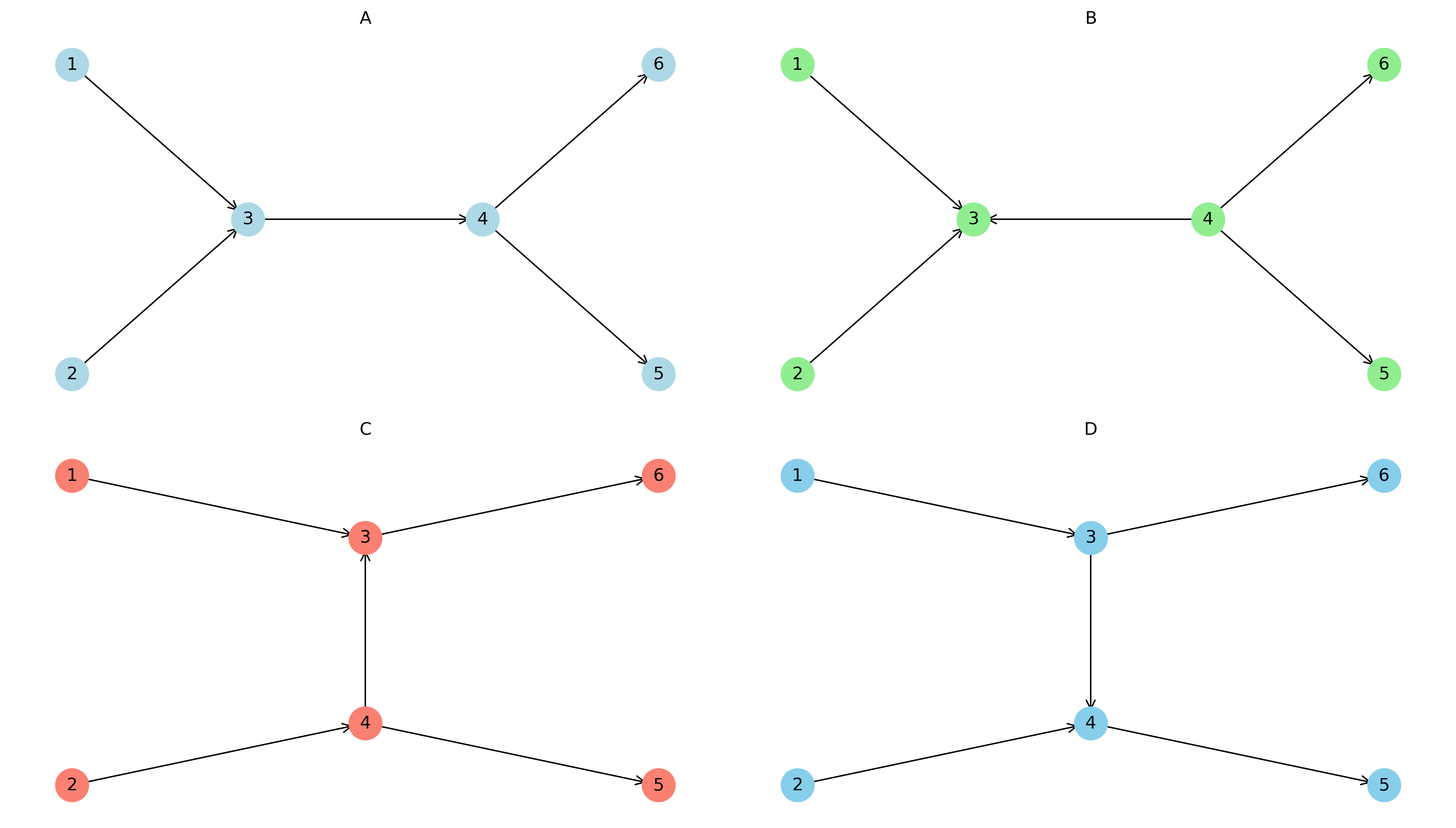}
    \caption{In the case of a Hamiltonian with three-wave interactions, the first-order contribution vanishes due to phase averaging of the interacting waves. However, the second-order terms yield a non-zero contribution, effectively resulting in a four-wave interaction mediated by a virtual (or forced) intermediate wave.}
    \label{fig:enter-label}
\end{figure}
 
 \begin{equation}
    \begin{aligned}
    \mathcal{H}_{3} & = \int d\mathbf{k}_1d\mathbf{k}_2d\mathbf{k}_3\bigg[G_{123}\left(\psi^*_1\psi_2\psi_3\delta(\mathbf{k}_1-\mathbf{k}_2-\mathbf{k}_3)+c.c\right)\bigg]\\
    \end{aligned}
\end{equation}
The 4-wave Hamiltonian will be given by the following expression.
\begin{equation}
    \begin{aligned}
    \mathcal{H}_{4} & = \int d\mathbf{k}_1d\mathbf{k}_2d \mathbf{k}_3\mathbf{k}_4  \bigg[W_{1234}\left(\psi_1\psi_2\psi^*_3\psi^*_4\delta(\mathbf{k}_1+\mathbf{k}_2-\mathbf{k}_3-\mathbf{k}_4)+c.c\right)\bigg]\\
    \end{aligned}
\end{equation}
The dynamical equations of the Hamiltonians can be derived using standard expressions, where the 3-wave Hamiltonians will give the following dynamics.
\begin{equation}
    \begin{aligned}
    i\frac{\partial\psi_\mathbf{k}}{\partial t} &= \omega_{\mathbf{k}}\psi_{\mathbf{k}}+\int\bigg[\frac{1}{2}V_{k12}\psi_{1}\psi_{2}\delta(\mathbf{k}-\mathbf{k}_1-\mathbf{k}_2)+V^*_{1k2}\psi_{1}\psi_{2}^*\delta(\mathbf{k}_1-\mathbf{k}-\mathbf{k}_2)\bigg]d\mathbf{k}_1d\mathbf{k}_2\\
    \end{aligned}
\end{equation}
The 4-wave Hamiltonian will have the following form.
\begin{equation}
    \begin{aligned}
    i\frac{\partial\psi_\mathbf{k}}{\partial t} &= \omega_{\mathbf{k}}\psi_{\mathbf{k}}+\frac{1}{4}\int\bigg[T_{k123}\psi_{1}^*\psi_{2}\psi_{3}\delta(\mathbf{k}+\mathbf{k}_1-\mathbf{k}_2-\mathbf{k}_3)\bigg]d\mathbf{k}_1d\mathbf{k}_2d\mathbf{k}_3\\
    \end{aligned}
\end{equation}
\section{Statistical approximation}
In Section 2, we derived the dynamic models governing the wave-like components that describe neural field activity. These waves are part of large-scale systems involving a multitude of distinct wave modes, where a statistical description is often necessary to capture the overall dynamics. In such systems, both the amplitude and phase of the $\mathbf{k}$-waves may vary. We will average over the phase to get a statistical variable of the neural field and neural field correlation functions, as defined in Equation~(15).
\begin{equation}
    \begin{aligned}
    \int  \psi_\mathbf{k}(\Omega) d\Omega & = \langle \psi_\mathbf{k}\rangle = 0\\
      \end{aligned}
\end{equation}
The correlations between wavefunctions to lowest order are shown in Equation~(16). 
\begin{equation}
    \begin{aligned}
    \langle \psi_\mathbf{k}\psi_\mathbf{k'}\rangle &= 0\\
    \langle \psi_\mathbf{k}\psi^*_\mathbf{k'}\rangle &= n(\mathbf{k})\delta(\mathbf{k}-\mathbf{k'})\\
    \langle \psi^*_\mathbf{k_1} \psi^*_\mathbf{k_2} \psi_\mathbf{k_3} \psi_\mathbf{k_4}\rangle &= n(\mathbf{k_1})n(\mathbf{k_2})\big[\delta(\mathbf{k_1}-\mathbf{k_3})\delta(\mathbf{k_2}-\mathbf{k_4})+\delta(\mathbf{k_1}-\mathbf{k_4})\delta(\mathbf{k_2}-\mathbf{k_3})\big]\\
      \end{aligned}
\end{equation}
The dynamics for $\mathbf{n}$ can be derived using Equation~(13) and its conjugate giving Equations~(17).
\begin{equation}
    \begin{aligned}
    i\frac{\partial\psi_\mathbf{k}}{\partial t} \psi_{\mathbf{k}}^{*}&= \omega_{\mathbf{k}}|\psi_{\mathbf{k}}|^2+\int\bigg[\frac{1}{2}V_{k12}\psi_{1}\psi_{2}\psi_{\mathbf{k}}^{*}\delta(\mathbf{k}-\mathbf{k}_1-\mathbf{k}_2)+V^*_{1k2}\psi_{1}\psi_{2}^*\psi_{\mathbf{k}}^{*}\delta(\mathbf{k}_1-\mathbf{k}-\mathbf{k}_2)\bigg]d\mathbf{k}_1d\mathbf{k}_2\\
    -i\frac{\partial\psi_\mathbf{k}^*}{\partial t} \psi_{\mathbf{k}}&= \omega_{\mathbf{k}}|\psi_{\mathbf{k}}|^2+\int\bigg[\frac{1}{2}V_{k12}^*\psi_{1}^*\psi_{2}^*\psi_{\mathbf{k}}\delta(\mathbf{k}-\mathbf{k}_1-\mathbf{k}_2)+V_{1k2}\psi_{1}^*\psi_{2}\psi_{\mathbf{k}}\delta(\mathbf{k}_1-\mathbf{k}-\mathbf{k}_2)\bigg]d\mathbf{k}_1d\mathbf{k}_2\\
    \end{aligned}
\end{equation}
Averaging over the phase, along with some intermediate derivations, yields the kinetic equations corresponding to the three-wave and four-wave Hamiltonians, given in Equations~(18) and (19), respectively.
\begin{equation}
    \begin{aligned}
    \frac{\partial n_\mathbf{k}}{\partial t} &= \pi\int\bigg[|V_{k12}|^2(n_1n_2-n_k(n_2+n_1))\delta(\mathbf{k}-\mathbf{k}_1-\mathbf{k}_2)\delta(\omega_k-\omega_1-\omega_2)\\
    &+2|V_{1k2}|^2(n_kn_2-n_1(n_2+n_k) )\delta(\mathbf{k}_1-\mathbf{k}-\mathbf{k}_2)\delta(\omega_1-\omega_k-\omega_2)\bigg]d\mathbf{k}_1d\mathbf{k}_2\\
  \end{aligned}
\end{equation}
\begin{equation}
    \begin{aligned}
    \frac{\partial n_\mathbf{k}}{\partial t} &=
    \frac{\pi}{2}\int\bigg[|T_{k123}|^2(n_2n_3(n_1+n_k)\\ & -n_1n_k(n_2+n_3))\delta(\mathbf{k}+\mathbf{k}_1-\mathbf{k}_2-\mathbf{k}_3)\delta(\omega_k+\omega_1-\omega_2-\omega_3)\bigg]d\mathbf{k}_1d\mathbf{k}_2d\mathbf{k}_3\\
  \end{aligned}
\end{equation}
These integrals will be denoted by $I(k, t)$, with the type of Hamiltonian (three-wave or four-wave) implicitly understood from the context.

\section{Wave Turbulence}
We now describe various theoretical possibilities for steady-state spectral features. Equations~(18) and~(19) characterise the effects of three-wave and four-wave interactions on the wave action, or equivalently, the flux of wave energy through $k$-space. The steady state is defined by the condition $\frac{\partial n}{\partial t} = 0$, indicating a constant spectral composition. Furthermore, in an equilibrium state, the wave action remains unchanged by definition. This condition allows for the derivation of several expressions that constrain the dynamics of the system, often referred to as constants of motion. We denote the total energy of the system by $E$, and the total wave action by $N$.
\begin{equation}
    \begin{aligned}
  E & = \int \omega n(\mathbf{k},t) d\mathbf{k}\\
  N &= \int n(\mathbf{k},t) d\mathbf{k}\\
    \end{aligned}
\end{equation}
These conserved qualities will have an accompanying conserved current, which will be given by the following expression.
\begin{equation}
    \begin{aligned}
  \frac{\partial n}{\partial t} +\nabla \cdot p & = 0 \\
    \end{aligned}
\end{equation}
The negative of the divergence of the current is given by the collision integral giving us the following expression.
\begin{equation}
    \begin{aligned}
  \frac{\partial n}{\partial t} & = I \\
    \end{aligned}
\end{equation}

\subsection{Equilibrium solutions}
Another conserved quantity is the entropy, as given in Equation~(23).

\begin{equation}
    \begin{aligned}
  S & = \int \ln n(\mathbf{k},t) d\mathbf{k}\\
    \end{aligned}
\end{equation}
Varying the entropy while keeping the energy constant yields the equilibrium solutions.

\begin{equation}
    \begin{aligned}
  \delta(S+\mu E) & = \delta\bigg[\int \ln n(\mathbf{k},t) d\mathbf{k}+\mu\int \omega n(\mathbf{k},t) d\mathbf{k}\bigg]\\
    \end{aligned}
\end{equation}
The equilibrium solution of the system is given in Equation~(25).
\begin{equation}
    \begin{aligned}
  0 & = \frac{1}{n(\mathbf{k},t)}+\mu\omega\\
  n(\mathbf{k},t) & = -\frac{1}{\mu\omega} = \frac{T}{\omega}
    \end{aligned}
\end{equation}
The temperature of the system has been defined using the Lagrange multiplier, $\mu$.

\subsection{Non-equilibrium solutions - Kolmogorov solutions}
The non-equilibria solutions with steady state dynamics also known as the the Kolmogorov solutions can be derived using the kinetic equation or the wave-action dynamics, equation~(18) and ~(19). These can be derived if the system is isotropic with scale invariance. We can define scale invariance as shown in Equation~(26). 
\begin{equation}
    \begin{aligned}
 V(\lambda k_1,\lambda k_2,\lambda k_3) & = \lambda^m V(k_1, k_2,k_3)\\
  \end{aligned}
\end{equation}
The energy flux will be given by Equation~(27).
\begin{equation}
    \begin{aligned}
  \frac{dP(k)}{dk} & =-(2k)^{d-1}\pi\omega(k)I(k)\\
  \end{aligned}
\end{equation}
This relation can be integrated over the k-variable. 
\begin{equation}
    \begin{aligned}
 P(k) & \simeq\int_0^{k} dk (2k)^{d-1}\pi\omega(k)\frac{\partial n}{\partial t}= \int_0^{k} dk (2k)^{d-1}\pi\omega(k)I(k)\\
\end{aligned}
\end{equation}
Equation~(28) can be used to estimate a power law distribution of $n$ over $k$ defined in Equation~(29). 
\begin{equation}
    \begin{aligned}
 n(k) & = Ak^{-s}\\
  \end{aligned}
\end{equation}
Using dimensional analysis, steady state solutions (or Kolmogorov states) can be estimated where for the integral, $I$, we have the dimensions in $k$ as follows:
\begin{equation}
    \begin{aligned}
        I & \approx \int |V|^2n^2\delta(k)\delta(\omega)dk^2\\
        &\approx k^{2m-2s-d-\alpha+2d}
  \end{aligned}
\end{equation}
We get the following dimensions for the energy flux, $P(k)$.
\begin{equation}
    \begin{aligned}
 P(k) & \simeq\int_0^{k} dk (2k)^{d-1}\pi\omega(k)\frac{\partial n}{\partial t} \\
 & \approx  k^{d+\alpha}k^{2m-2s+d-\alpha}\\
 & = k^{2(m+d-s)}
\end{aligned}
\end{equation}
The wave action flux, $Q(k)$, will give a similar equation. 
\begin{equation}
    \begin{aligned}
 Q(k) & \simeq\int_0^{k} dk (2k)^{d-1}\pi\frac{\partial n}{\partial t}= \int_0^{k} dk (2k)^{d-1}\pi I(k)\\
\end{aligned}
\end{equation}
Dimensional analysis will give a similar relation as equation~(31).
\begin{equation}
    \begin{aligned}
 Q(k) & \simeq k^dk^{2m-2s+d-\alpha}=k^{2m-2s+2d-\alpha}\\
\end{aligned}
\end{equation}
Assuming a constant flux of energy or wave action (i.e., independent of $\mathbf{k}$) leads to steady-state solutions. The corresponding spectral exponents are denoted by $s_e$ for energy and $s_n$ for wave action.
\begin{equation}
    \begin{aligned}
        s_{n} & = s_{e}-\frac{\alpha}{2}\\
  \end{aligned}
\end{equation}
The distribution over the wave-frequency $\omega$ can be estimated using Equation~(34) and the dispersion relation.
\begin{equation}
    \begin{aligned}
 n_e(\omega) & = Ak^{-\alpha \frac{s_e}{\alpha}} = A\omega^{-\frac{m+d}{\alpha}}\\
 n_n(\omega) & =  A\omega^{-\frac{m+d-\frac{\alpha}{2}}{\alpha}} = A\omega^{-\frac{m+d}{\alpha}+\frac{1}{2}}\\
  \end{aligned}
\end{equation}
For the four-wave interaction, a similar set of relations can be derived. However, the collision integral $I$ will have a different dimensionality compared to the three-wave interaction case.
\begin{equation}
    \begin{aligned}
        I & \approx \int |V|^2n^3\delta(k)\delta(\omega)dk^3\\
        &\approx k^{2m+2d-3s-\alpha}
  \end{aligned}
\end{equation}
The flux of the energy and wave action are given in Equation~(37).
\begin{equation}
    \begin{aligned}
 P(k) & \approx  k^{d+\alpha}k^{2m+2d-3s-\alpha}\\
 & = k^{2m+3d-3s}\\
 Q(k) & \approx  k^{d}k^{2m+2d-3s-\alpha}\\
 & = k^{2m+3d-3s-\alpha}\\
\end{aligned}
\end{equation}
The distribution over the wave frequency, $\omega$, is given in Equation~(38).
\begin{equation}
    \begin{aligned}
 n_e(\omega) & =  A\omega^{-\frac{\frac{2m}{3}+d}{\alpha}}\\
 n_n(\omega) & =  A\omega^{-\frac{\frac{2m}{3}+d}{\alpha}+\frac{1}{3}}\\
  \end{aligned}
\end{equation}

\subsection{Single and dual cascade solutions of spectral decay}
The steady states described above correspond to cascades of either energy or wave action. These systems are often referred to as single-cascade systems, as the full spectral distribution within the inertial range is governed by the cascade of a single conserved quantity across different regions of $k$-space. Systems exhibiting dual cascades, where two conserved quantities cascade simultaneously, have been described by \citet{kraichnan1967inertial}.

Building on the results derived in Section~4.2, we can construct a dual-cascade model. Part of the cascade corresponds to an energy cascade and the other to a wave action cascade. We have observed that the difference in spectral gradient between these regimes is either $\frac{1}{2}$ or $\frac{1}{3}$, depending on whether the system is governed by three-wave or four-wave interactions.

From energy conservation principles, we expect an inverse wave action cascade from a given input scale $k_0$ toward smaller wavenumbers, and a direct energy cascade toward larger wavenumbers, where the energy is dissipated. A characteristic increase in spectral gradient is observed in dual-cascade systems in fluid dynamics, where the inverse cascade is of energy and the direct cascade involves enstrophy (vorticity squared), transferring from small to large $k$ \cite{yakhot1993hidden,she1993universal}. In such systems, the energy input scale corresponds to the transition point in the spectral slope.

Theoretical estimates of the spectral gradients for these steady-state systems can be found in \cite{boffetta2007energy,kraichnan1967inertial}. Unlike fluid dynamics, wave systems lack a direct analogue to vorticity in their governing equations. However, it is possible to define a variable in wave dynamics that exhibits dynamical characteristics analogous to vorticity in fluid systems.

\subsubsection{Vortices in wave theory}
Enstrophy arises from the conservation of vorticity in the dynamics. In fluid turbulence, the vorticity equation is obtained by taking the curl of the Navier–Stokes equation, which governs the dynamics of the fluid; see Equation~(39).
\begin{equation}
    \begin{aligned}
 \frac{\partial}{\partial t}\mathbf{u} +   \mathbf{u}\cdot\nabla\mathbf{u} & = -\frac{1}{\rho}\nabla p + \nu\nabla^2\mathbf{u} \\
 \boldsymbol{\omega} & = \nabla \times \mathbf{u}\\
    \end{aligned}
\end{equation}
In the 2D case we get a simplification of Equation~(39). 
\begin{equation}
    \begin{aligned}
 \frac{\partial}{\partial t} \boldsymbol{\omega} +\boldsymbol{\omega}\cdot\nabla  \mathbf{u} & = \nu\nabla^2\boldsymbol{\omega} \\
 \end{aligned}
\end{equation}
The dynamical equations governing neural fields differ fundamentally from those in fluid dynamics and do not contain an intrinsic vorticity term. However, it can be shown that wave dynamics governed by complex fields—i.e., waves characterised by distinct amplitude and phase variables—can be transformed into a fluid-like dynamical framework. A commonly used complex dynamical equation in the modelling of neural fields is the nonlinear Schrödinger equation, which captures key features of wave propagation and interaction in such systems.
\begin{equation}
    \begin{aligned}
        i\partial_{t} \phi & = \partial_{i}^{2}\phi + |\phi|^2\phi \\
\end{aligned}
\end{equation}
This equation can be transformed into a fluid-like dynamical form using the Madelung transformation, as demonstrated in \cite{spiegel1980fluid}. By expressing the fields of the nonlinear Schrödinger equation in terms of amplitude and phase variables, and performing the appropriate differentiations, one obtains a dynamical equation closely resembling the Navier–Stokes equation. 
\begin{equation}
    \begin{aligned}
        \phi & = \sqrt{\rho} e^{i\psi}\\
\end{aligned}
\end{equation}
In this formulation, the fluid density is defined as $\rho$, and the velocity field is given by $\mathbf{u} = 2\nabla \psi$, where $\psi$ denotes the phase of the complex wave function. Importantly, the velocity arises from the gradient of the phase variable, highlighting the correspondence between wave phase dynamics and fluid motion.
\begin{equation}
    \begin{aligned}
\frac{Du}{Dt} = \partial_{t} \mathbf{u} +(\mathbf{u}\cdot\nabla)\mathbf{u}& = -\frac{\nabla \rho^2}{\rho}+2\nabla\frac{\nabla^2\sqrt{\rho}}{\sqrt{\rho}}\\
\frac{\partial\rho}{\partial t} + \nabla\cdot(\rho \mathbf{u}) &= 0
\end{aligned}
\end{equation}
The resulting equation corresponds to a variation of Euler flow for compressible fluids—specifically, an isentropic compressible Euler flow with an adiabatic index of 2 \cite{nazarenko2011wave}. The final term on the right-hand side is often referred to as the \textit{quantum pressure} term, due to its origin in the Schrödinger equation. Taking the curl of this equation yields the vorticity equation, thereby linking the phase structure of wave dynamics to rotational features in fluid-like systems.
\begin{equation}
    \begin{aligned}
\frac{D\boldsymbol{\omega}}{Dt} &= (\boldsymbol{\omega}\cdot\nabla)\mathbf{u}-\boldsymbol{\omega}(\nabla\mathbf{u})+ \frac{\nabla\rho\times\nabla\rho^2}{\rho^2}\\
\end{aligned}
\end{equation}
Similarly, the Klein–Gordon field, including the connectivity gauge, can be transformed into a hydrodynamical form exhibiting vorticity \cite{takabayasi1953remarks,cooray2024cortical}. The spectral index for the vorticity cascade can be estimated using a relation from \cite{kraichnan1967inertial}. Specifically, the vorticity cascade index $a_{\omega}$ is related to the energy cascade index $a_{e}$ by the following expression:
\begin{equation}
    \begin{aligned}
 a_\omega & =  \frac{1}{2}(3a_e+1)\\
  \end{aligned}
\end{equation}

\subsection{Multifractal state}
The steady-state solutions described above are based on the Richardson cascade \citep{kolmogorov1991local,frisch1995turbulence}, in which eddies transfer energy to smaller eddies, each occupying an equal volume of space. The spatial dimension of these eddies is typically 2 or 3, depending on whether the fluid flows on a plane or within a volume.

In multifractal models of turbulence \citep{frisch1995turbulence}, the energy-containing structures do not fill the entire spatial domain but are instead supported on fractal sets characterized by a non-integer Hausdorff dimension. The Hausdorff dimension is a measure of the fractal geometry of a set, quantifying how its detail changes with scale and capturing its complex, self-similar structure.

A similar fractal dynamics can be assumed for wave turbulence, where the family of interacting eddies or waves occupies a fractal subset of the $k$-space. This fractal support alters the spectral indices of the Kolmogorov-type solutions, as demonstrated in Equation~(46).

\begin{equation}
    \begin{aligned}
 a_e & = \frac{m+2d-D}{\alpha}\\
 a_n & =  \frac{m+2d-D}{\alpha}-\frac{1}{2}\\
  \end{aligned}
\end{equation}
The dimension of the fractal set to which the dynamics of the interactions are attracted is denoted by $D$. When this dimension equals the full spatial dimension $d$, the cascade dynamics reproduce the original Kolmogorov spectra. As $D$ decreases towards zero, the spectral slope becomes progressively steeper.

It is also possible for the dynamics to be attracted to different fractal structures simultaneously. In such a \emph{multifractal} state, characterized by a spectrum of fractal dimensions, the Kolmogorov spectra are modified so that the constant power-law decay may either increase or decrease as $k$ tends to zero or infinity \cite{frisch1995turbulence,mandelbrot1974intermittent,benzi1984multifractal,sreenivasan1997phenomenology}.

\subsection{Wave turbulence and steady state spectra} 
Experimental recordings of cortical activity have been performed in various animal models as well as in humans. In humans, such recordings are routinely used in the clinical assessment of cortical function, primarily for the diagnosis and monitoring of epilepsy. Recordings are obtained either non-invasively, using sensors placed on the scalp (electroencephalography, EEG), or invasively, with electrodes placed intracranially—either on the cortical surface (subdural electrodes) or within the cortical tissue itself (stereo-EEG).

All of these recordings exhibit a characteristic frequency spectrum featuring an inverse power-law decay of the form $\frac{1}{f^{a}}$ over approximately the 1–100 Hz range, often referred to as the inertial region in wave turbulence terminology. This power-law component is also known as the aperiodic component of the EEG \cite{donoghue2022methodological}. The exponent $a$ has been estimated in humans to lie roughly between 2 and 3. This steady-state spectral feature is observed across different brain states, including both wakefulness and sleep.

Experimental data also reveal changes in spectral features at higher frequencies (above 100–200 Hz), where the spectrum either decays rapidly or saturates at a minimal level. Figure~2 illustrates a schematic overview of the different spectral patterns observed in cortical tissue recordings.

\begin{figure}
    \centering
    \includegraphics[width=0.75\linewidth]{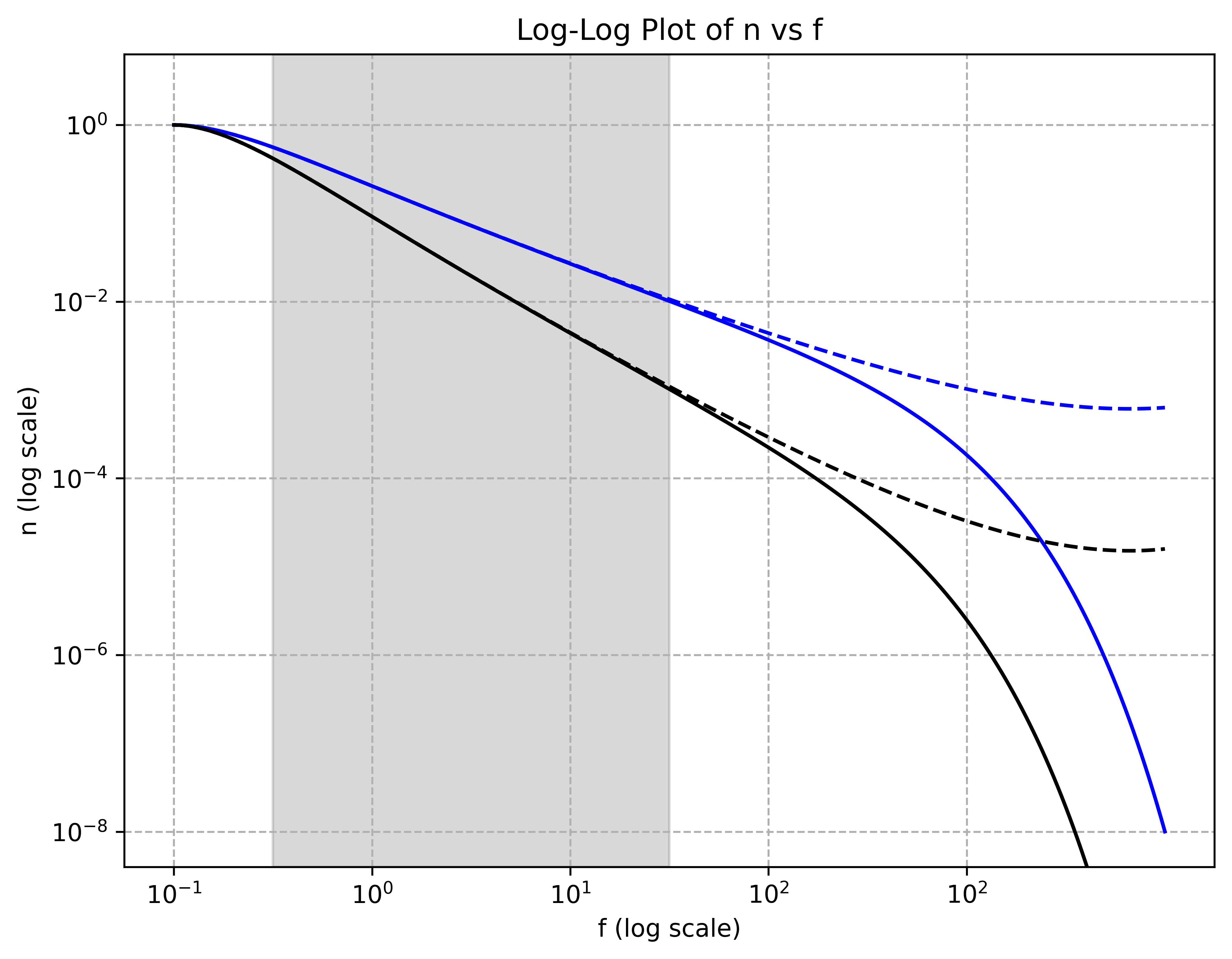}
    \caption{Log-log plot of the squared amplitude versus frequency observed in cortical activity. Within the shaded grey region, the spectral features exhibit a linear decay with a slope varying between 2 and 3. This grey region, corresponding to the inertial range of the wave dynamics, spans approximately from 1 to 100 Hz. At higher frequencies, the spectral data either decay rapidly (solid lines) or saturate to a constant level (dashed lines). At lower frequencies (to the left of the grey region), the spectral features also exhibit saturation.}
    \label{fig:enter-label}
\end{figure}

Figure 3 shows empirical estimates of the spectral features, highlighting a central inertial region with a characteristic power-law decay. Outside this inertial range, the spectral features saturate as frequencies increase or decrease.
 \begin{figure}
    \centering
    \includegraphics[width=0.75\linewidth]{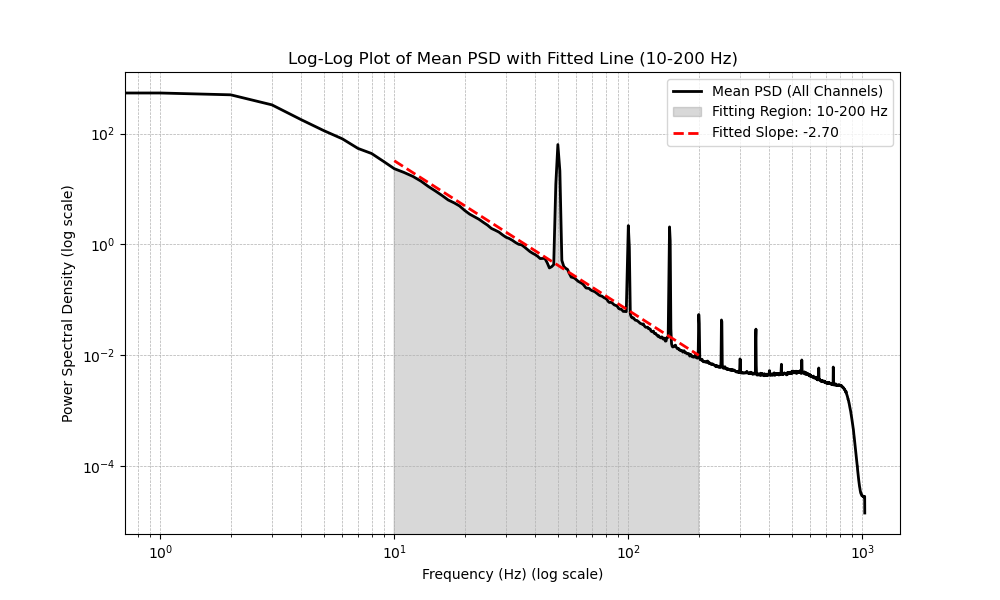}
    \caption{Spectral features of cortical activity recorded using intracranial electrodes. The central region (shaded in grey) exhibits a power-law decay with an estimated slope of approximately $-2.7$. Outside this region, the spectrum saturates, followed by a sharp decay near 1000 Hz. Note the presence of artifactual spikes at 50 Hz and its harmonics.}
    \label{fig:enter-label}
\end{figure}
Assuming a frequency power-law decay with an exponent between 2 and 3, we can estimate certain features of the underlying model. Considering the classical Kolmogorov spectra (i.e., without fractal subdimensions), the following relations hold for three-wave interactions:
\begin{equation}
    \begin{aligned}
 a_e & = \frac{m+d}{\alpha}\approx \frac{m+2}{1}\\
 \end{aligned}
\end{equation}

We have assumed an approximately linear relation between the wave frequency and momentum ($\alpha \approx 1$). The scaling parameter $m$ is difficult to define precisely due to uncertainty about which model best describes the overall dynamical behavior of neural fields. However, it can be argued that $m$ may be estimated using the above relations. In fluid dynamic turbulence, where the underlying dynamical equations are better understood, a value of $m = 0.5$ is commonly used. This corresponds to a power-law decay exponent of approximately $-2.5$, consistent with the empirical observations shown in Figure~3. For four-wave interactions, a slightly smaller exponent is predicted according to Equation~(46).
\begin{equation}
    \begin{aligned}
 a_e & = -\frac{\frac{2m}{3}+d}{\alpha} \approx -\frac{\frac{2m}{3}+2}{1} \\
 \end{aligned}
\end{equation}
The above equations involve two unknown variables: the scaling factor $m$ and the exponent $\alpha$. The multifractal dynamics described in Section~4.4 yield a similar expression but include an additional variable representing the fractal dimension of the structure containing the eddies. In general, the resulting steady-state spectra are steeper.

The multifractal model also accounts for the saturation of the wave action $n$ at low frequencies (or small $k$ values), as well as the rapid decay or saturation observed at high $k$ values. 

All scenarios discussed so far in this section involve a single cascade within the inertial range. However, several studies have reported spectral features consistent with a dual cascade. Specifically, two distinct regions with different power-law decay slopes have been observed \citep{miller2009power}. The decay exponent at high frequencies was approximately $-4$, whereas for smaller $k$ values it was around $-2.5$. 

Equation~(45) predicts a decay exponent of $-4$ for the second cascade if the first energy cascade has a decay exponent of $-2.5$, consistent with the findings illustrated in Figure~4.
\begin{figure}
    \centering
    \includegraphics[width=0.75\linewidth]{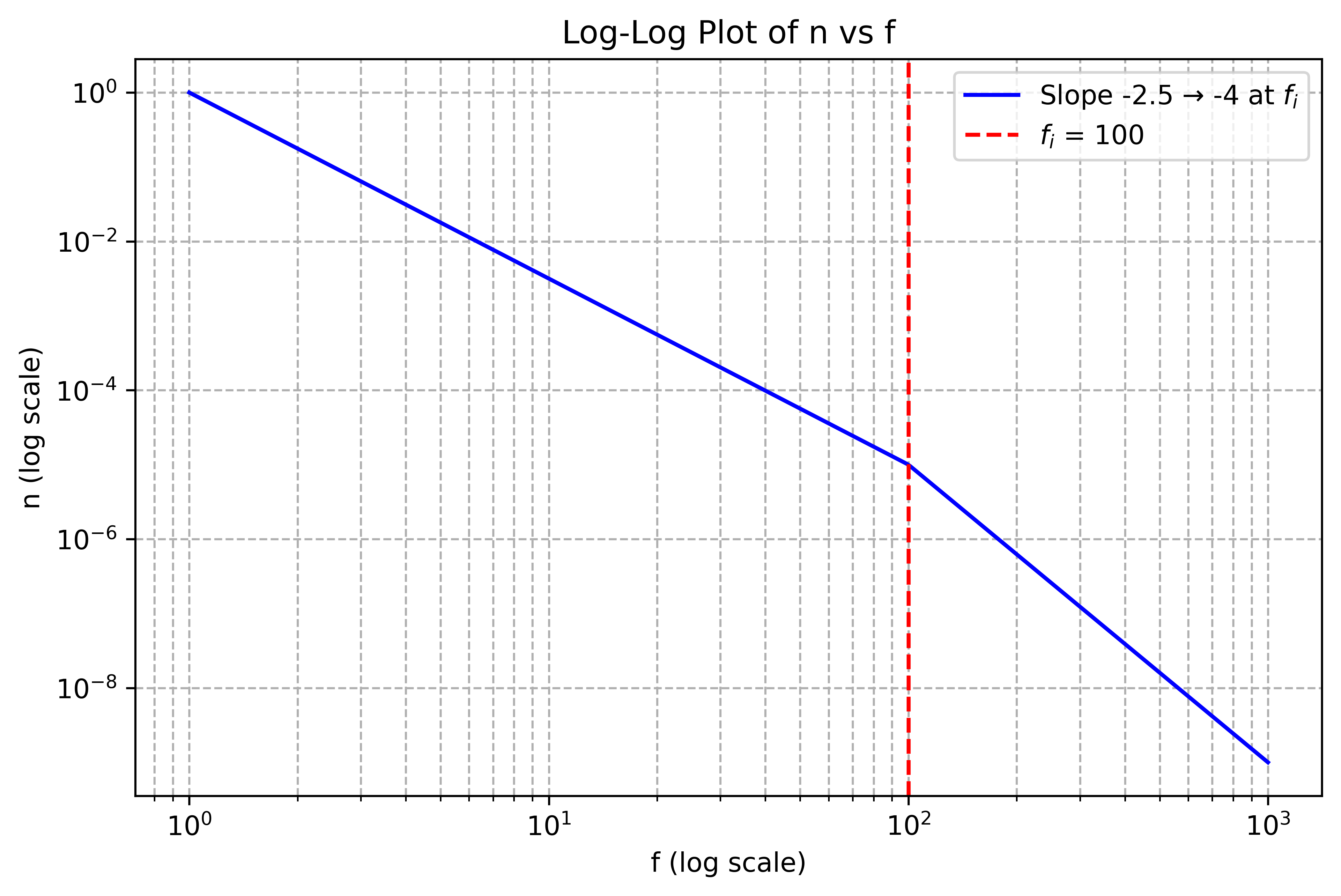}
    \caption{Double cascade system with an inverse energy cascade below 100 Hz with a decay exponent of approximately $-2.5$. Direct cascade is seen above 100 Hz involving vorticity transfer to higher frequencies. Empirical data support these characteristic decay values.}
    \label{fig:enter-label}
\end{figure}

\section{Sources and Sinks of the Steady State}
The Kolmogorov solution, which assumes a constant energy flux, requires both a source and a sink to sustain a non-zero flux. These are typically located far apart in $k$-space. In regions sufficiently distant from the source and sink, there is no significant energy creation or dissipation, and the system remains in a conservative steady state.

In this section, we investigate how the presence of sources and sinks modifies the Kolmogorov steady state. The spectral characteristics of EEG recordings exhibit several features that can inform the nature of the underlying dynamics. We will examine the effects of both wideband and narrowband sources, as well as the impact of high-frequency damping on steady-state solutions. The source/sink term, denoted by $\Gamma$, modifies the governing expressions developed in Sections~3 and~4.
\begin{equation}
    \begin{aligned}
 \frac{\partial n(k,t)}{\partial t} & =  I(n_k)+n_k\Gamma(k)\\
  \end{aligned}
\end{equation}
Note that the Kolmogorov spectrum retains the same power-law decay rate within the inertial range, where the source/sink term $\Gamma$ is assumed to be zero. In this regime, the energy flux is conserved, and the spectral shape remains unaffected by the presence of sources or sinks located outside the inertial range. In equation~50, $C$ is a constant.
\begin{equation}
    \begin{aligned}
 n(k,t) & =  C P^{\frac{1}{2}}k^{-(m+d)} \\
  \end{aligned}
\end{equation}

\subsection{Wide band Sources}
It can be shown that a wide band source will not affect the power law gradient upto a constant logarithmical term \cite{zakharov2012kolmogorov}. 

\subsection{Narrow band Sources}
A narrowband source can be represented mathematically as a Dirac delta function, $\delta(k_0)$, centred at the injection scale $k_0$.
\begin{equation}
    \begin{aligned}
        \Gamma(k,t) & = \Gamma_0\delta(k_0)\\
  \end{aligned}
\end{equation}
The effect of a narrowband source on the spectrum can be estimated qualitatively and will depend on the specific form of nonlinearity in the underlying Hamiltonian of the system. For a three-wave interaction nonlinearity, the result is a cascade of spectral spikes superimposed on the underlying Kolmogorov spectrum. Let $n_1(k)$ denote the original Kolmogorov spectrum and $n_2(k)$ the source-induced modulation. The total spectrum is then given by the combination of these two components, where the presence of $n_2(k)$ introduces localised spikes at harmonics of the source frequency.
\begin{equation}
    \begin{aligned}
        n(k,t) & = n_1(k,t) + n_2(k,t)\\
  \end{aligned}
\end{equation}
The nonlinear interactions between the source-induced spikes generate recurrent spectral features at integer multiples of the injection scale $k_0$, i.e., at $k = n k_0$ for integer $n$.
\begin{equation}
    \begin{aligned}
        n_2(k,t) & =\sum_j a_j\delta(jk_0) \\
  \end{aligned}
\end{equation}
As a first approximation, it can be assumed that the activity (or amplitude) within each spike follows the same decay as the Kolmogorov spectrum.
\begin{equation}
    \begin{aligned}
        n_2(k,t) & =\sum_j j^{-(m+d)}k_0^{-(m+d)}\delta(jk_0) \\
  \end{aligned}
\end{equation}
However, each spectral peak will exhibit a finite width, denoted by $\Delta k_0$, which increases with each successive cascade of the spikes. This broadening can be understood intuitively, as the system involves a two-wave to one-wave interaction process, leading to increasing spectral dispersion at higher harmonics.

\begin{equation}
    \begin{aligned}
        k_0 \pm \delta k \rightarrow 2k_0 \pm 2\delta k\\
  \end{aligned}
\end{equation}
Each spike can be estimated as a triangular region.
\begin{equation}
    \begin{aligned}
        \delta(k_0) \rightarrow \Delta(k_0,\delta k)\\
  \end{aligned}
\end{equation}
The interactions will lead to a cascading set of spikes.
\begin{equation}
    \begin{aligned}
        n_2(k,t) & =\sum_j j^{-(m+d)-1}k_0^{-(m+d)}\Delta(jk_0,j\delta k) \\
  \end{aligned}
\end{equation}
The full spectra will be given in Equation~(58). 
\begin{equation}
    \begin{aligned}
        n(k,t) & = \lambda P^{\frac{1}{2}}k^{-(m+d)}  + \sum_j j^{-(m+d)-1}k_0^{-(m+d)}\Delta(jk_0,j\delta k) \\
  \end{aligned}
\end{equation}
Experimental and numerical simulations have provided evidence supporting a cascade of spectral spikes consistent with three-wave interaction dynamics (see Figure 5; \citep{sheremet2019wave}). In contrast, four-wave interactions result primarily in the dispersion and broadening of spectral peaks over time, a phenomenon that has also been observed in some datasets \citep{hirai1999enhanced}.

\begin{figure}
    \centering
    \includegraphics[width=0.75\linewidth]{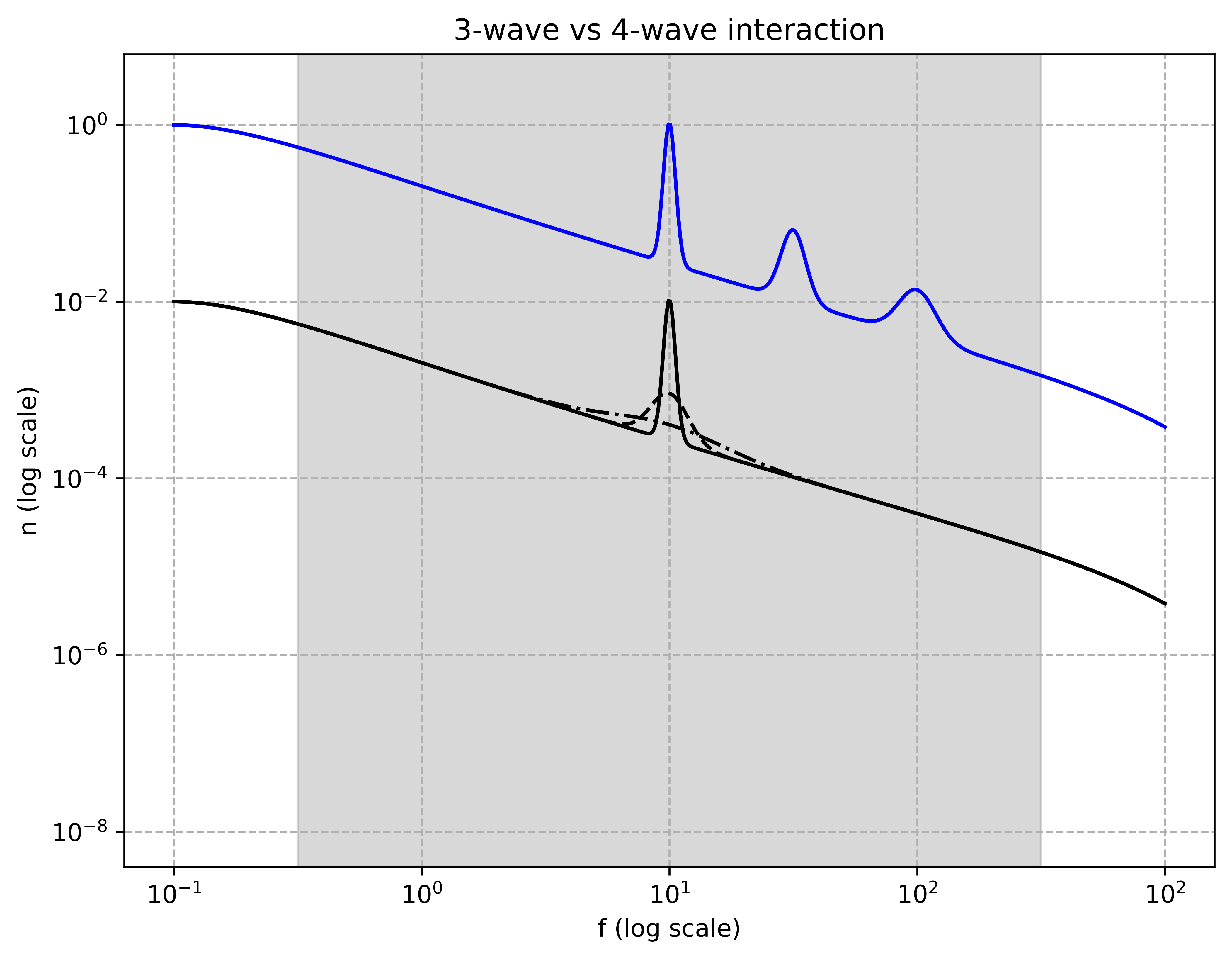}
    \caption{Narrow spectral pumping in three-wave (blue) and four-wave (black) turbulence interactions. In the three-wave interaction regime, harmonic peaks emerge at integer multiples of the pumping frequency. These harmonics are generated through a process in which two waves within the pumped region combine to form a new wave with approximately twice the frequency. This interaction recurs iteratively, with higher-order harmonics arising from successive interactions between previously generated peaks. In contrast, four-wave interactions exhibit a fundamentally different behaviour: two waves within the pumped region scatter off each other, generating wave pairs with a range of frequency ratios. This results in a progressive temporal dispersion of the pumped spectral activity rather than the formation of discrete harmonics.}
    \label{fig:enter-label}
\end{figure}

\subsection{High freuqency sinks}
In a similar way the analysis of a sink or dissipation of the waves can be analysed where the function $\Gamma$ will be negative for large $k$. Pertubative analysis and numerical stidues have shows that the existence of a sharp dissipiation at a given value of $k$ there will be a reduction in the power law deacy followed by an exponential drop in the spectrum \cite{zakharov2012kolmogorov}. This is due to the creation of a bottle neck in k-space causing the flux of waves to reduce close to the point of sharp dissipation.

\subsection{Affect of transients on cortical activity}
In the preceding sections, we examined the effects of continuous sources on cortical activity. In this section, we investigate the impact of short-lived, transient sources on the dynamics of cortical activity. The partial differential equation that governs the system under these conditions is given by Equation~(59).

\begin{equation}
    \begin{aligned}
 \frac{\partial n_k}{\partial t} & =  I(n_k)+n_k\Gamma(k,t)\\
  \end{aligned}
\end{equation}
A Dirac spike will be used to model a short initial source.
\begin{equation}
    \begin{aligned}
\Gamma(k,t) & = \delta(t-t_0)\delta(k-k_0)\\
  \end{aligned}
\end{equation}
This yields the following expression (Equations~(61)), neglecting the first term on the right-hand side.
\begin{equation}
    \begin{aligned}
        \frac{\partial n_k}{\partial t} & =  n_k\delta(t-t_0)\delta(k-k_0)\\
        \frac{1}{n_k}\frac{\partial n_k}{\partial t} & =  \delta(t-t_0)\delta(k-k_0)\\
        \ln n_k & =  \Theta(t-t_0)\delta(k-k_0)\\
    \end{aligned}
\end{equation}
The resulting waveform will be concentrated around the site of energy insertion.
\begin{equation}
    \begin{aligned}
        n(k_0,t)>>n(k,t) \\
    \end{aligned}
\end{equation}
There is a significant elevation in wave action, $n$, at the site of energy introduction compared to other regions in $k$-space. However, the collision integral encompasses dynamical processes that allow for the redistribution of this introduced activity across $k$-space. In the previous section, we described how 3-wave interactions can generate higher-order spectral spikes at integer multiples of $k_0$. One aspect not previously discussed is the role of disintegrative interactions, where a single wave splits into two daughter waves with reduced momentum and energy. As there is no strict constraint on the energy ratio between these resulting waves, this mechanism allows for a diffusion of spectral energy toward lower $k$ values—particularly on the lower-frequency (left) side of the induced spike. The corresponding collision integral governing this process is provided in Equation~(63).
\begin{equation}
    \begin{aligned}
    I(n_\mathbf{k}) &= \pi\int\bigg[|V_{k12}|^2(n_1n_2-n_k(n_2+n_1))\delta(\mathbf{k}-\mathbf{k}_1-\mathbf{k}_2)\delta(\omega_k-\omega_1-\omega_2)\\
    &+2|V_{1k2}|^2(n_kn_2-n_1(n_2+n_k) )\delta(\mathbf{k}_1-\mathbf{k}-\mathbf{k}_2)\delta(\omega_1-\omega_k-\omega_2)\bigg]d\mathbf{k}_1d\mathbf{k}_2\\
  \end{aligned}
\end{equation}
To investigate the dynamics on the lower-$k$ (left) side of the spectral spike, we consider the scattering processes in which waves redistribute into slightly smaller $k$ values. This corresponds to interactions that result in the transfer of energy or wave action from the vicinity of the spike towards lower frequencies.
\begin{equation}
    \begin{aligned}
        1 & \rightarrow k + 2\\
        n_1 & = n_k+\delta n \\
        n_2 & = \delta n \\
    \end{aligned}
\end{equation}
Expanding the collision integral given in equation~(63), we obtain equation~(65).
\begin{equation}
    \begin{aligned}
    I(n_\mathbf{k}) &= \pi\int\bigg[|V_{k12}|^2(n_1n_2-n_k(n_2+n_1))\delta(\mathbf{k}-\mathbf{k}_1-\mathbf{k}_2)\delta(\omega_k-\omega_1-\omega_2)\\
    &+2|V_{1k2}|^2(n_kn_2-n_1(n_2+n_k) )\delta(\mathbf{k}_1-\mathbf{k}-\mathbf{k}_2)\delta(\omega_1-\omega_k-\omega_2)\bigg]d\mathbf{k}_1d\mathbf{k}_2\\
    & \approx \pi\int\bigg[|V_{112}|^2(-n_1^2)\delta(\mathbf{k}-\mathbf{k}_1-\mathbf{k}_2)\delta(\omega_k-\omega_1-\omega_2)\\
    &+2|V_{112}|^2(-n_1^2 )\delta(\mathbf{k}_1-\mathbf{k}-\mathbf{k}_2)\delta(\omega_1-\omega_k-\omega_2)\bigg]d\mathbf{k}_1d\mathbf{k}_2\\
    \approx 2|V_{112}|^2n_k^2\\
  \end{aligned}
\end{equation}
With further simplification, we can estimate how the amplitude evolves on the lower-\(k\) side of the spike.
\begin{equation}
    \begin{aligned}
        \frac{\partial n_k}{\partial t} & = Cn^2_k\\
        \frac{dn_k}{n_k^2} & = C dt \\
        A-\frac{1}{Ct} & = n_k\\
    \end{aligned}
\end{equation}
We observe an increase in wave activity to the left of the spike, indicating that this part of the spectral deformation results in a deflection of the spike towards lower \(k\)-values. To estimate the evolution of small perturbations in this region, we employ the linearised form of the collision integral \cite{zakharov2012kolmogorov}.
\begin{equation}
    \begin{aligned}
        \frac{\partial \delta n_k}{\partial t} &=\int d\mathbf{k}_1d\mathbf{k}_2|V_{kk_1k_2}|^2\delta(\mathbf{k}-\mathbf{k}_1-\mathbf{k}_2)
        \delta(\omega_{k}-\omega_{1}-\omega_{2})([n_2-n_k]\delta n_1\\
        &+[n_1-n_k]\delta n_2-[n_1+n_2]\delta n_k)\\
        &-|V_{1k2}|^2\delta (\mathbf{k}_1-\mathbf{k}-\mathbf{k}_2)\delta(\omega_{1}-\omega_{k}-\omega_{2})([n_k-n_1]\delta n_2\\
        &+[n_k+n_2]\delta n_1-[n_2+n_1]\delta n_k)\\
    \end{aligned}
\end{equation}
We then obtain the following expression at the left edge of the perturbation from the steady state:
\begin{equation}
    \begin{aligned}
        \frac{\partial \delta n_k}{\partial t} &=-\int d\mathbf{k}_1d\mathbf{k}_2|V_{1k2}|^2\delta (\mathbf{k}_1-\mathbf{k}-\mathbf{k}_2)\delta(\omega_{1}-\omega_{k}-\omega_{2})([n_k-n_1]\delta n_2\\
        &+[n_k+n_2]\delta n_1-[n_2+n_1]\delta n_k)\\
        & \approx -|V_{1k2}|^2 n_2\frac{\partial \delta n_k}{{\partial k}}
    \end{aligned}
\end{equation}
This equation describes a wave traveling to the left with speed $c=|V_{1k2}|^2 n_2$. We assume $n_2$ remains constant, as it is small and situated in a region of the wave far from the perturbative bump on the left. Depending on the dispersion relation, this leads to a reduction in frequency as detailed below.

A comparison with observed frequency changes during electrographic seizure activity on the cortical surface reveals a similar temporal pattern. The modelled evolution of seizure frequency over time follows this trend, as shown in Figure~6 \cite{lagarde2019repertoire}.

\begin{equation}
    \begin{aligned}
        \omega_k & = A(k_0-|V_{1k2}|^2 n_2t)^\alpha\\
        \frac{\partial \omega}{\partial t} & = -\alpha A|V_{1k2}|^2 n_2(k_0-|V_{1k2}|^2 n_2t)^{\alpha-1}\\
    \end{aligned}
\end{equation}

\begin{figure}
    \centering
    \includegraphics[width=0.75\linewidth]{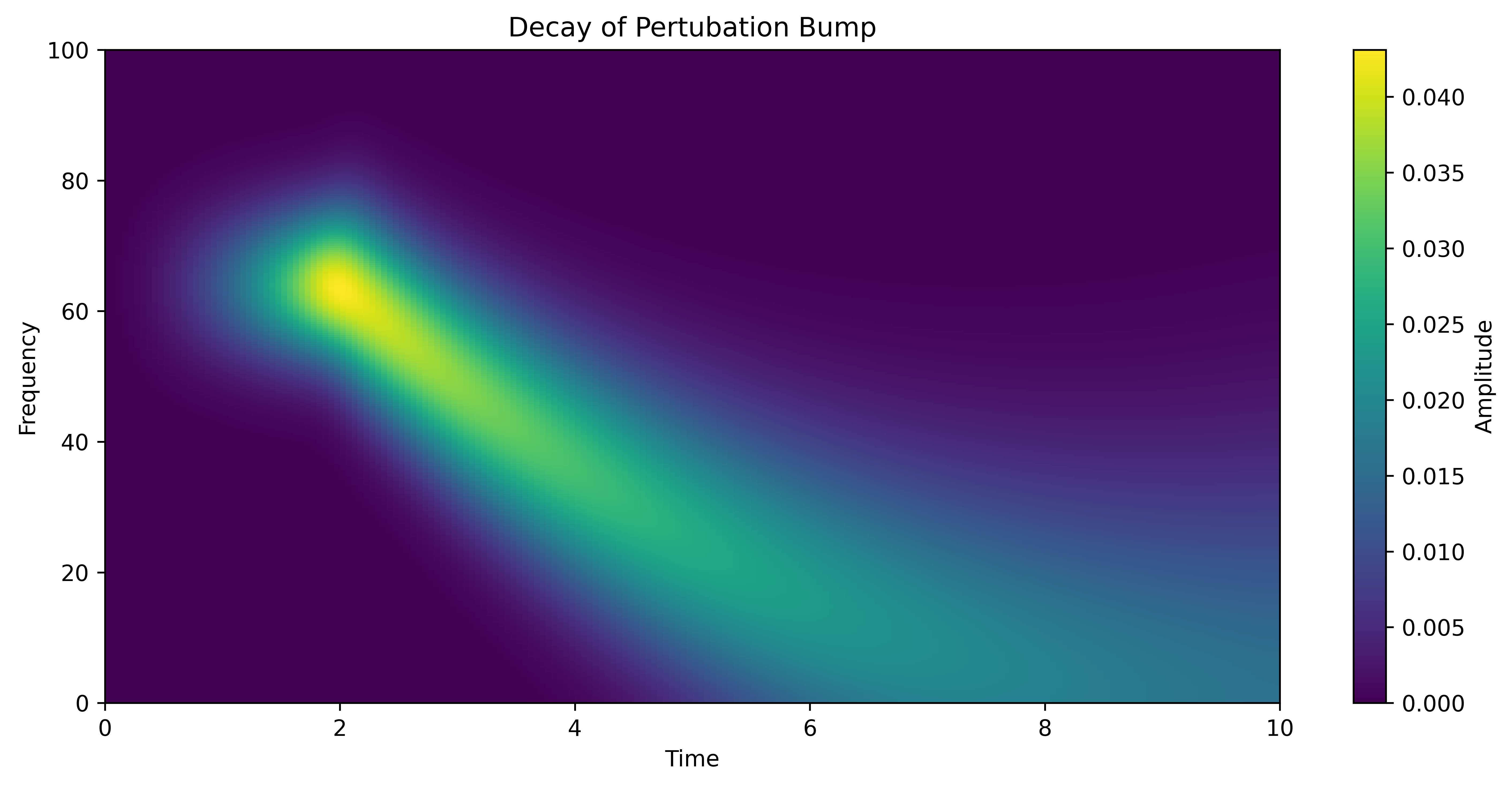}
    \caption{Estimation of the decay of a perturbative bump around the steady-state solution reveals a power-law decay in frequency, accompanied by dispersion of the signal. The time–frequency plot of EEG activity changes during seizure onset exhibits a similar pattern, often referred to as high-frequency, low-amplitude seizure onset activity.}
    \label{fig:enter-label}
\end{figure}
\section{Discussion}
In this paper, we discussed the theoretical foundation of wave turbulence and its relevance to the analysis of neural fields. We derived key statistical features of the underlying dynamics, including the characteristic power-law decay that is consistently observed in recordings from cortical tissue. A range of possible spectral patterns was presented, along with the theoretical mechanisms that may generate them. These theoretical predictions were compared with representative examples from the literature, highlighting matching spectral features observed in empirical data.

A comprehensive analytical treatment of turbulence was first developed by Kolmogorov in 1941, building on the experimental observations available at the time. For fluid turbulence, a power-law decay of energy with respect to wavenumber was predicted, characterised by an exponent of $\frac{5}{3}$. This result has been extensively verified in various experimental settings; for a comprehensive introduction and further references, see \cite{frisch1995turbulence}. 

Interestingly, a similar power-law decay is a robust feature of recordings from cortical tissue, where the observed spectral exponent is typically higher, estimated to lie between 2 and 3. While a direct link between classical fluid dynamics and the dynamics of cortical tissue is not immediately evident, there has long been a suggestion that the brain supports wave-like interactions. This idea is foundational to the theory of wave turbulence, notably described by Zakharov \cite{zakharov2012kolmogorov}. The precise spectral decay rate in wave turbulence depends on the specific interaction model, and it has been proposed that models of cortical dynamics might be constrained or identified by matching their predicted decay rates to those empirically observed.

However, the range of theoretical models capable of producing spectral decays within the empirically observed range is quite broad. Furthermore, the spectral patterns observed in neural recordings are more intricate than the classical turbulence treated by Kolmogorov. The presence of chaotic or fractal attractors in the dynamics introduces additional parameters — such as the Hausdorff dimension — that further influence the decay rate. In fact, most nonlinear models can be tuned to produce either classical Kolmogorov spectra or multifractal decays \cite{benzi1984multifractal,meneveau1991multifractal}.

The literature on recordings from cortical tissue is notably variable. Several differences can be attributed to recording methodologies, as the size and configuration of the recording electrode significantly influence the data. However, when local field potentials (LFPs) or macroelectrodes sample activity from a sufficiently large neural population, the dynamics tend to be oscillatory rather than spiking. Spectral analysis of such oscillatory signals often reveals a power-law decay of amplitude with frequency. Some studies report a single power-law regime, while others identify two distinct regimes — a phenomenon consistent with either a single energy cascade or a dual cascade in turbulent systems \cite{deco2020turbulent,miller2009power}.

Moreover, cortical spectral activity frequently exhibits a combination of aperiodic (power-law) and periodic features. From the perspective of wave turbulence, periodic features could be interpreted as spectral pumping. The consequences of such pumping differ across studies: in some cases, the pumped activity disperses; in others, it generates harmonics. This discrepancy may reflect differences in the underlying wave interaction mechanisms. Specifically, systems governed by 3-wave interactions are known to produce harmonics, whereas 4-wave interactions tend to cause dispersion of energy introduced via pumping.

Determining the underlying interaction model is challenging, as empirical features point to multiple theoretical frameworks. For example, the nonlinear Schrödinger equation (NLSE) represents a canonical 4-wave interaction system. However, it is known that systems governed by the NLSE can undergo dynamical transitions—akin to Bose-Einstein condensation—into effective 3-wave systems \cite{pitaevskii1961vortex}. Currently, the available spectral data are insufficient to conclusively identify the governing dynamics of cortical fields.

Given that there is still no complete analytical understanding of turbulence in fluid or wave systems—despite decades of study and well-defined first-principles models—it is overly optimistic to expect a full derivation of neural field dynamics solely from spectral data. Nevertheless, certain features must be incorporated into any plausible model. These include the capacity for both 3-wave and 4-wave interactions, and the potential for either single or dual cascade phenomena, as observed in empirical data.

\section{Conclusion} 
This paper explored cortical dynamics through the framework of wave turbulence. By drawing analogies with fluid turbulence, we examined how spectral features such as power-law decay, energy cascades, and non-linear interactions may arise from wave-based neural activity.

Empirical recordings from cortical tissue often show power-law spectral decay and transient harmonic features, which suggest turbulent-like processes. We discussed how 3-wave and 4-wave interaction models lead to distinct spectral signatures—harmonic generation and dispersion, respectively—and how these may explain patterns seen during events like electrographic seizures.

While a complete analytical model of neural turbulence remains out of reach, the observed features constrain the space of viable models. Future work should aim to further link theoretical dynamics with measurable spectral characteristics, potentially revealing deeper insights into the principles governing large-scale brain activity.

\backmatter



\section*{Declarations}
\subsection*{Funding}
No funding was received for the paper. 
\subsection*{Conflict of Interest/Competing Interests}
The author (GC) does not have competing interests to declare that are relevant to the content of this article.

\bibliography{sn-bibliography}

\end{document}